\documentclass[
    ,final            % use final for the camera ready runs
  ]
  {aipproc}

\layoutstyle{6x9}
\usepackage{amsfonts}
\usepackage{amssymb}
\begin{document}
\newcommand{\beqn}{\begin{eqnarray}}
\newcommand{\eeqn}{\end{eqnarray}}
\newcommand{\be}{\begin{equation}}
\newcommand{\ee}{\end{equation}}
\newcommand{\non}{\nonumber \\}
\newcommand{\om}{\overline{m}}
\newcommand{\vol}{{\cal V}}
\newcommand{\tom}{{\tilde \omega}}
\newcommand{\st}{{Stueckelberg~}}
\newcommand{\te}{\theta}
\newcommand{\thb}{\bar\theta}
\newcommand{\vp}{\varphi}
\newcommand{\mathsym}[1]{{}}
\newcommand{\unicode}{{}}

\title{The Stueckelberg Extension  and\\ Milli Weak and Milli Charged Dark Matter}

%<Replace this text with PACS numbers; choose from this list:
%               \texttt{}}
%     \texttt{http://www.aip..org/pacs/index.html}>}

\author{Daniel Feldman, Zuowei Liu and Pran Nath }{address={Department of Physics, Northeastern University, Boston, MA 02115, USA
}}

%\author{<author2>}{
%  address={<common address for author2 and author3>}
%}

%\author{<author3>}{
%  address={<common address for author2 and author3>}
%  ,altaddress={<author1 address>} % additional visiting address
%}

\begin{abstract}
A overview is given of the recent developments in the $U(1)_X$
Stueckelberg extensions of the Standard Model and of MSSM where all
the Standard Model particles are neutral under the $U(1)_X$,  but an
axion  which is absorbed  is charged under both $U(1)_X$ and
$U(1)_Y$ and acts as the connector field coupling the Standard Model
sector with the \st sector. Coupled with the usual Higgs mechanism
that breaks the $SU(2)_L\times U(1)_Y$ gauge symmetry, this scenario
produces mixings in  the neutral gauge  boson sector generating an
extra $Z'$ boson. The  couplings of the extra $Z'$ to the Standard
Model particles are milli weak but its couplings to the hidden
sector matter, defined as matter that couples only to the gauge
field of  $U(1)_X$,  can be of normal electro-weak strength.  It is
shown that such extensions, aside from the possibility of leading to
a sharp $Z' $ resonance, lead to two new types of dark matter: milli
weak (or extra weak) and milli charged. An analysis of the relic
density shows that the WMAP-3 constraints can be satisfied for
either of these scenarios.  The types of models discussed could
arise as possible field point limit of certain Type IIB orientifold
string models.

\end{abstract}

\maketitle

%%%%%%%%%%%%%%%%%%%%%%%%%%%%%%%%%%%%%%%%%%%%
%% MAINMATTER
%%%%%%%%%%%%%%%%%%%%%%%%%%%%%%%%%%%%%%%%%%%%
\section{Introduction}
Through a Stueckelberg mechanism an Abelian gauge boson develops
mass without the benefit of a Higgs mechanism (For the early history
of the Stueckelberg mechanism see,
\cite{stueckelberg:38,ogievetskii:62, Chodos:1971yj,Kalb:1974yc}).
Thus consider the Lagrangian \beqn {\cal L}_0 = -\frac14 {\cal
F}_{\mu\nu}{\cal F}^{\mu\nu} - \frac12 (m A_\mu + \partial_\mu
\sigma)(m A^\mu + \partial^\mu \sigma)\ , \eeqn
which is gauge invariant under the transformations $\delta A_\mu =
\partial_\mu \lambda \ , \quad \delta \sigma = - m \lambda$. With
the gauge fixing term ${\cal L}_{\rm gf} = -  \left( \partial_\mu
A^\mu + \xi m \sigma \right)^2/2\xi$, the total Lagrangian reads
\beqn {\cal L}= -\frac14 {\cal F}_{\mu\nu}{\cal F}^{\mu\nu} -
\frac{m^2}{2} A_\mu A^\mu - \frac{1}{2\xi} (\partial_\mu A^\mu)^2
 - \frac12 \partial_\mu \sigma \partial^\mu \sigma - \xi
\frac{m^2}{2} \sigma^2 +  g A_\mu J^\mu, \eeqn
where we have added also an interaction term which contains the
coupling of $A_{\mu}$ with fermions via a conserved current with
$\partial_\mu J^\mu = 0$.  Here the fields $\sigma$ and $A_{\mu}$
are decoupled  and renormalizability and unitarity are manifest.
Mass growth by the Stueckelberg mechanism occur quite  naturally  D
brane constructions where one  encounters  the group $U(N)$ for a
stack of $N$ D branes which is then broken to its  subgroup $SU(N)$
via Stueckelberg couplings. Thus, for example, one has
\beqn
\stackrel{\rm D ~branes} {U(3) \times U(2) \times U(1)^2}
~~\stackrel{\rm Stueckelberg}{\longrightarrow}~~ \stackrel{\rm
SM}{SU(3)_C\times SU(2)_L \times U(1)_Y}\ . \eeqn

\section{The \st extension of SM} 
The \st extension can be used for the extensions of the Standard Model \cite{Kors:2004dx}
and of MSSM \cite{Kors:2004ri,Kors:2005uz,Kors:2004iz}. We begin by discussing the \st
extension of the  Standard Model \cite{Kors:2004dx} where we write the Lagrangian so that
${\cal L}_{\rm StSM} = {\cal L}_{\rm SM}+{\cal L}_{\rm St}$,
where
\beqn
{\cal L}_{\rm St} = -\frac{1}{4} C_{\mu\nu}C^{\mu\nu}
+ g_X C_\mu J^\mu_X
- \frac{1}{2} (\partial_{\mu}\sigma  + M_1 C_{\mu} + M_2 B_{\mu})^2\ .
\eeqn
It is easily checked that the above  Lagrangian is invariant under
the following transformations :
$\delta_{Y}(C_{\mu},B_{\mu},\sigma)=(0,\partial_{\mu}\lambda_Y,-M_2\lambda_Y)$
and
$\delta_{X}(C_{\mu},B_{\mu},\sigma)=(\partial_{\mu}\lambda_X,0,-M_1\lambda_X)$.
The two Abelian gauge bosons can be decoupled from $\sigma$ by the
addition of gauge fixing terms as before. Additionally, of course,
one has to add the standard gauge fixing terms for the SM gauge
bosons to decouple from the Higgs.

We look now at the physical content of  the theory. In the vector
boson sector in the basis $V_{\mu}^T = ( C_{\mu}, B_{\mu}, A_{\mu}^3
)$,  the mass matrix for the vector bosons takes the form
\begin{equation} M^{2}_{[V]}=
 \left[\matrix{
    M_1^2             & M_1M_2                                & 0 \cr
    M_1M_2         &M_2^2 + \frac{1}{4}v^2g_Y^2  & -\frac{1}{4}v^2g_2g_Y \cr
      0                   & -\frac{1}{4}v^2g_2g_Y            & \frac{1}{4}v^2g_2^2
}\right], \label{stmass}
 \end{equation}
where $g_2$ and $g_Y$ are the $SU(2)_L$ and $U(1)_Y$ gauge coupling
constants, and are normalized so that $M_W^2=g_2^2v^2/4$. It is
easily checked that  $\det(M^{2}_{[V]} )=0$ which implies that one
of the eigenvalues is zero, whose eigenvector we identify with the
photon. The remaining two eigenvalues are non-vanishing and
correspond to the $Z $ and $Z'$ bosons.  The symmetric matrix
$M^{2}_{[V]} $ can be diagonalized by an orthogonal transformation,
$V={\cal O}E$, with $E_{\mu}^T = ( Z'_{\mu}, Z_{\mu}, A_{\mu}^{\gamma}
)$ so that the eigenvalues are given by the set : $ {\rm diag(
M^{2}_{[V]}})=\{ M^2_{{\rm Z}'} , M^2_{\rm Z} , 0 \} $. One can
solve for ${\cal O}$ explicitly and we use the parametrization
\beqn %\label{A}
{\cal O}= \left[\matrix{ \cos\psi \cos\phi
-\sin\theta\sin\phi\sin\psi & -\sin\psi \cos\phi
-\sin\theta\sin\phi\cos\psi & -\cos\theta \sin\phi\cr \cos\psi
\sin\phi +\sin\theta\cos\phi\sin\psi & -\sin\psi \sin\phi
+\sin\theta\cos\phi\cos\psi & \cos\theta \cos\phi\cr
-\cos\theta\sin\psi & -\cos\theta\cos\psi & \sin\theta }\right],
%\ .
\nonumber
\eeqn
where  $\tan (\phi) = \frac{M_2}{M_1} \equiv \epsilon
\ , \quad \tan (\theta) ~=~ \frac{g_Y}{g_2}\cos(\phi) ~=~
\tan(\theta_W)\cos(\phi) \ .$
 The third angle is given by
$\tan (\psi) = {\tan(\theta)\tan(\phi)M_{{\rm W}}^2}/
                     ({\cos(\theta)(M_{{\rm Z}'}^2-M_{\rm W}^2(1+\tan^2(\theta)))})$. 
This allows one to choose $\epsilon$ and $M_1$ as two independent
parameters to characterize physics beyond SM.  There is also  a
modification of the expression of the electric charge in terms of SM
parameters. Thus if we write the EM interaction in the form $e
A_\mu^\gamma J^\mu_{\rm em}$ the expression for $e$ is given by
\beqn e = {g_2g_Y\cos(\phi)}/{\sqrt{g_2^2+g_Y^2 \cos^2(\phi)}} \
. \eeqn The LEP and Tevatron data  puts stringent bounds on
$\epsilon$. One finds \cite{Feldman:2006ce,Feldman:2006wb} that it
is constrained by $\epsilon \lesssim .06$ in most of  the parameter
space. In the absence of a hidden sector, i.e., the matter sector
that couples only to $C_{\mu}$, the $Z'$ can decay only into visible
sector quarks and leptons, and its decay width is governed by
$\epsilon$ and hence the $Z'$ is very sharp, with a width that lies
in the range of of maximally several hundred MeV compared to several
GeV that one expects for a $Z'$ arising from a GUT group (a narrow
$Z'$ can also arise in other models, see e.g.,
\cite{Chang:2006fp,Battaglia:2005zf,Burdman:2006gy,Ferroglia:2006mj,Davoudiasl:1999jd}).
However, even a very sharp $Z'$ is discernible at the Tevatron and
at the LHC using the dilepton signal. On the  other hand if a hidden
sector exists with normal size gauge coupling to the $C_{\mu}$ then
$Z'$ can decay into the hidden sector particles and will have a
width in the several GeV range. In this case the  branching ratio of
$Z'$ to $l^+l^-$ will be very small
\cite{Cheung:2007ut,Feldman:2007wj} and the dilepton signal will not
be detectable. We will return to this issue in the context of milli
charged dark matter.

\section{\st extension of the minimal supersymmetric  standard model
}

To obtain the supersymmetric Steuckelberg extension \cite{Kors:2004ri,Kors:2004iz,Kors:2005uz}
 we consider the
 Stueckelberg chiral multiplet
$S=(\rho+i\sigma,\chi,F_S)$ along with the vector superfield multiplets for
the $U(1)_Y$ denoted by $B=(B_\mu,\lambda_B,D_B)$
and for the $U(1)_X$ denoted by  $C=(C_\mu,\lambda_C,D_C)$.
The \st addition to the SM Lagrangian is then given by
\beqn {\cal L}_{\rm St} = \int d^2\te d^2\thb\ (M_1C+M_2B+  S +\bar
S )^2. \label{mass} \eeqn
Under $U(1)_Y$ and $U(1)_X$ the supersymmetrized gauge
transformations are then given by: $\delta_Y (C, B, S) =(0,
\Lambda_Y + \bar\Lambda_Y,- M_2 \Lambda_Y)$ and $\delta_X
(C,B,S)=(\Lambda_X + \bar\Lambda_X,0,- M_1 \Lambda_X)$.
Expanding the fields in the component form, in the Wess-Zumino
gauge, we have for a vector superfield, denoted here by $V=(C,B)$,
\beqn V~=~ -\theta\sigma^{\mu}\bar \theta V_{\mu} +i\theta\theta
\bar\theta \bar \lambda_V -i\bar\theta\bar\theta \theta  \lambda_V
+\frac{1}{2} \theta \theta\bar\theta\bar\theta D_V\ . \eeqn The
superfield $S$ in component notation is given by
 \beqn \label{superS} S &=& \frac{1}{2}(\rho
+i\sigma ) + \theta \chi
 + i \theta\sigma^{\mu}\bar\theta \frac{1}{2}(\partial_{\mu} \rho
+i \partial_{\mu} \sigma) \nonumber\\
&& + \theta\theta F_S + \frac{i}{2} \theta \theta \bar\theta
\bar\sigma^{\mu} \partial_{\mu}\chi
+\frac{1}{8}\theta\theta\bar\theta\bar\theta (\Box \rho+i\Box
\sigma)\ . \eeqn

We note that the superfield S  contains  the scalar
$\rho$ and the axionic pseudo-scalar $\sigma$.   In component form
${\cal L}_{\rm St}$ then has the form \beqn \label{stueck} {\cal
L}_{\rm St} &=& - \frac{1}{2}(M_1C_{\mu} +M_2 B_{\mu}
+\partial_{\mu} \sigma)^2
 - \frac{1}{2} (\partial_\mu \rho)^2
- i \chi \sigma^{\mu} \partial_{\mu}\bar \chi +2|F_S|^2
\\
&&  %\hspace{-1.2cm}
 +\rho(M_1D_C +M_2 D_B)
 +\big [ \chi (M_1 \lambda_C + M_2 \lambda_B)
 + {\rm h.c.} \big]\ .
\nonumber
\label{ls}
\eeqn
To the above we can add the gauge fields of the Standard Model which give

\beqn \hspace{-.5cm}
{\cal L}_{\rm gkin} &=&
-\frac{1}{4} C_{\mu\nu} C^{\mu\nu} -\frac{1}{4} B_{\mu\nu} B^{\mu\nu} -
i \lambda_B\sigma^{\mu}\partial_{\mu} \bar \lambda_B
-i \lambda_C\sigma^{\mu}\partial_{\mu} \bar \lambda_C
+\frac{1}{2} D_C^2  +\frac{1}{2} D_B^2\ .
\nonumber
\eeqn
The gauge fields can be coupled to the chiral superfields $\Phi_i$
of matter in the  usual way
\beqn
{\cal L}_{\rm matt} ~=~ \int d^2\te d^2\thb\, \Big[
\sum_i \bar \Phi_i e^{2g_Y Q_Y B+ 2g_X Q_X C} \Phi_i
 + \sum_i \bar \Phi_{{\rm hid},i} e^{2g_Y Q_Y B+ 2g_X Q_X C} \Phi_{{\rm hid},i}\Big] \ .
\nonumber
\eeqn
Here $Q_Y=Y/2$, and where $Y$ is the hypercharge so that
$Q=T_3+Y/2$. We assume that the SM matter fields do not carry any
charge under the hidden gauge group, i.e. $Q_X \Phi_i =0$. The \st
extensions of the  type we have discussed could have origin in Type
IIB orientifold models
\cite{Ghilencea:2002da,Ghilencea:2002by,Ibanez:2001nd,Antoniadis:2002qm,Blumenhagen:2002vp} and
several recent works appear to recover in its low energy limit the
type of models discussed here
\cite{Anastasopoulos:2006da,Anastasopoulos:2006cz,Coriano:2005js,Coriano:2007xg,Coriano:2007fw,
Anastasopoulos:2007qm,Coriano:2006rg}.

\subsection{Milli weak dark matter in $U(1)_X$ extension}
We note that the \st extension brings in two more  Majorana spinors
which we can construct out of the Weyl spinors as follows
$
\psi_S^T~=~(\chi_{\alpha},  \bar \chi^{\dot{\alpha}} ),~~~~
\lambda_X^T~=~
(\lambda_{C\alpha},  \bar \lambda^{\dot{\alpha}}_C)$.
This enlarges the neutralino mass matrix from being $4\times 4$ as is the case
in MSSM to a $6\times 6$ mass matrix in the \st extension.
The enlarged neutralino mass matrix reads
\beqn \label{neutrmass} M_{1/2} = \left[\matrix{ 0 & M_1 & M_2 & 0 &
0 & 0\cr M_1& \tilde m_X & 0 & 0 & 0 & 0\cr M_2& 0 & \tilde m_1 & 0
& -c_{\beta}s_{W}M_0 & s_{\beta}s_WM_0\cr 0 & 0 & 0 & \tilde m_2 &
c_{\beta}c_{W}M_0 & -s_{\beta}c_WM_0 \cr 0 & 0 & -c_{\beta}s_{W}M_0
&  c_{\beta}c_{W}M_0 & 0 & -\mu \cr 0 & 0 & s_{\beta}s_{W}M_0  &
-s_{\beta}c_{W}M_0 &  -\mu & 0}\right] . \eeqn
Here  the $4\times 4$ matrix on the lower right hand corner  is the
usual neutralino mass matrix of MSSM, while the $2\times 2$ matrix
in the top left hand corner is due the \st extension. The term
$\tilde m_X$ is the soft breaking term which is added by hand.   The
zero entry in the  upper left hand corner arises due to the Weyl
fermions not acquiring soft masses.  The $6\times 6$ matrix gives
rise to six Majorana mass eigenstates which may be  labeled as
follows $ E_{[1/2]}=(\chi_1^0,\chi_2^0,\chi_3^0,\chi_4^0,\chi_5^0,
\chi_6^0)^T$, where the two additional Majorana eigenstates
$(\chi_5^0, \chi_6^0)$ are due to the Stueckelberg extension. We
label these two $\xi_1^0, \xi_2^0$ and to leading order in
$\epsilon$ their masses are given by
\beqn m_{\xi_1^0}\simeq \sqrt{M^{2} +\frac{1}{4}\tilde m_X^{2}}
-\frac{1}{2} \tilde m_X\ ,\quad m_{\xi_2^0}\simeq\sqrt{M^{2}
+\frac{1}{4}\tilde m_X^{2}} +\frac{1}{2} \tilde m_X\ . \eeqn
where $M^2=M_1^2+M_2^2$. 
If the mass of $\xi_1^0$ is less than the mass of other sparticles,
then $\xi_1^0$ will be a candidate for dark matter with R parity
conservation. These are  what one may call XWIMPS  (mWIMPS)
for extra (milli) weakly interacting massive particles.
Here the satisfaction of relic density requires  coannihilation
and one has to consider processes of the type
$\xi^0+\xi^0 \to  X\ , \quad
\xi^0+\chi^0   \to   X'\ , \quad
\chi^0+\chi^0  \to   X''\ $,
 where $\{X\}$ etc denote the
Standard Model final states.
In this case we can write the effective cross section as follows\cite{Feldman:2006wd}
 \beqn \sigma_{\rm eff}=
\sigma_{\chi^0\chi^0}     \frac{1}{(1+Q)^2}  (Q
+\frac{\sigma_{\xi^0\chi^0} }{\sigma_{\chi^0\chi^0} })^2\ ,
~~Q= \frac{g_{\chi^0}}{g_{\xi^0}} (1+\Delta)^{\frac{3}{2}}
e^{-x_f\Delta}\ . \eeqn
Here $g$  is the degeneracy for the corresponding particle, $x_f=m_{\xi^0}/T_f$ where
$T_f$ is the freeze-out temperature,
 and   $\Delta
=(m_{\chi^0}-m_{\xi^0})/m_{\xi^0}$ is the mass gap.
For the case of  XWIMPS one has
${\sigma_{\xi^0\chi^0} }/{ \sigma_{\chi^0\chi^0} } \sim {\cal
O}(\epsilon^2)\ll 1$.
  Now it is easily seen that when the mass  gap between $\xi^0$ and
$\chi^0$ is large and $x_f\Delta \gg 1$,  then $\sigma_{\rm eff}$ is
much  smaller than the typical WIMP cross-section and in this case
one does not have an efficient annihilation of the XWIMPS. On the
other hand if the mass gap between the XWIMP and WIMP is small then
coannihilation  of  XWIMPs is efficient.  In this case $Q\sim 1$ and
one has $\sigma_{\rm eff}\simeq \sigma_{\chi^0\chi^0}
\left(\frac{Q}{1+Q}\right)^2 $. The  above result is valid more
generally with  many channels participating in the coannihilations,
as can be seen by defining an effective Q given by
  $Q=\sum_{i=2}^{N}
Q_i$ where $Q_i=(g_i/g_1) (1+\Delta_i)^{3/2}e^{-x_f \Delta_i}$.
Thus, satisfaction of the relic density constraints arise quite
easily for the XWIMPS. A detailed analysis of the relic density of
XWIMPS was carried out in \cite{Feldman:2006wd} and it was found
that the WMAP-3 constraint\cite{Spergel:2006hy} 
$ \Omega_{CDM}
h^2 =0.1045^{+0.0072}_{-0.0095}$ can be satisfied by XWIMPS.

\section{\st mechanism with kinetic mixing}
We discuss now the \st extension with kinetic mixing \cite{Feldman:2007wj} for which we
take the Lagrangian to be of the form
$\mathcal{L}_{\rm StkSM} = \mathcal{L}_{\rm SM} +\Delta  \mathcal{ L}$ where
\begin{eqnarray}
\Delta\mathcal{L} & \supset &  -\frac{1}{4} C_{\mu \nu}C^{\mu \nu}
- \frac{\delta}{2}C_{\mu\nu}B^{\mu \nu}
 -\frac{1}{2}(\partial_{\mu}\sigma+ M_1 C_{\mu}+M_2 B_{\mu})^2
+ g_{X}J^{\mu}_{X}C_{\mu}.
\label{stksm}
\end{eqnarray}
In this case the kinetic mixing matrix,in the basis $V^T=(C,B,A^3)$
is,
\begin{equation}
 \mathcal{K}= \left[\matrix{
 1 & \delta &0  \cr
    \delta & 1 &0  \cr
    0 & 0 & 1
}\right].
\end{equation}
A simultaneous diagonalization of the kinetic energy and of the mass matrix can
 be obtained by a transformation $T=KR$, which is a combination of a $GL(3)$ transformation ($K$)
 and an orthogonal transformation ($R$). This allows one to work in the diagonal basis,
 denoted by $E^T=(Z',Z,A^{\gamma})$, through the  transformation  $V =(K R) E$,
where the matrix $K$ which diagonalizes  the kinetic terms has the
form
\begin{equation}
 K = \left[\matrix{
 C_{\delta} & 0 &0 \cr
    -S_{\delta} & 1 &0 \cr
    0 & 0 & 1
}\right],\hspace{.15 cm}
C_{\delta}=\frac{1}{\sqrt{1-\delta^2}},\hspace{.15 cm}
S_{\delta}=\delta C_{\delta}.
\label{kinetic}
\end{equation}
The diagonalization also
leads to the following relation for the  electronic charge
\begin{equation}
\frac{1}{e^2}=\frac{1}{g_2^2}+\frac{1-2\epsilon\delta+\epsilon^2}{g_Y^2}.
\label{e}
\end{equation}
Thus  $g_Y$ is related to $g_Y^{SM}$ by $ g_{Y}= \gamma
\sqrt{1+\epsilon^2 -2 \delta \epsilon},\hspace{.5cm} \gamma \equiv
g^{SM}_Y$. In the absence of a hidden sector, there is only  one
parameter that enters in the analysis of electroweak fits. This
effective parameter is given by $\bar \epsilon =(\epsilon-
\delta)/\sqrt{1-\delta^2}$. Thus one can satisfy the LEP and the
Tevatron electro-weak data with $\bar\epsilon \lesssim .06$ but
$\epsilon$ and $\delta$ could be individually larger.

\begin{figure}
  \includegraphics[height=.3\textheight]{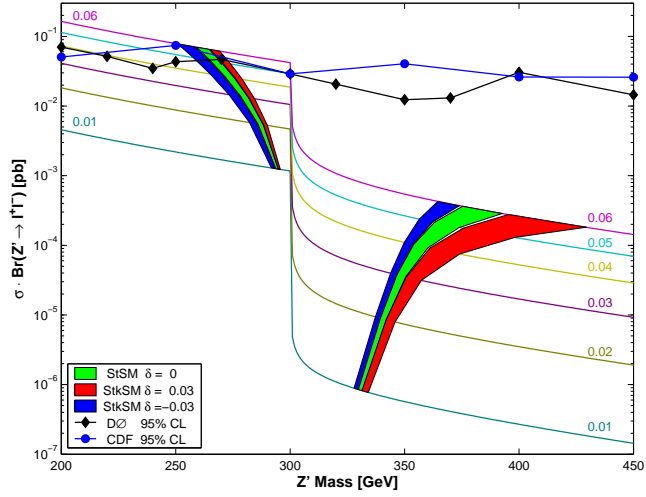}
  \caption{
  The colored regions indicate the satisfaction of the relic density constraints consistent
  with the WMAP-3 constraints
  and the size of the  dilepton signal $\sigma \cdot
Br(Z'\rightarrow l^+l^-)$ at the Tevatron
 as a function of  $M_{Z'}$ when
$2M_{\chi}=300$ GeV. The curves in ascending order are for values of
$\bar{\epsilon}$ in the range $(0.01-0.06)$ in steps of 0.01. The
dilepton signal has a dramatic fall as $M_{Z'}$ crosses the point
$2M_{\chi}=300$ GeV where the $Z'$ decay into the hidden sector
fermions is kinematically allowed, widening enormously the $Z'$
decay width. The green shaded regions are where the WMAP-3 relic
density constraints are satisfied for the case when there is no
kinetic mixing. Red and blue regions are for the case when kinetic
mixing is included. The current constraints on the dilepton and
signal from CDF\cite{CDF} and the D\O\ search for narrow resonances
\cite{Abazov:2005pi} are also exhibited. From \cite{Feldman:2007wj}.
}

\label{fig1}
\end{figure}

\subsection{How  milli charge is generated in \st extension   }

To exhibit the phenomenon of generation of milli-charge in the \st model we consider
two gauge fields $A_{1\mu}, A_{2\mu}$ corresponding to the gauge  groups $U(1)$ and $U(1)'$.
We choose  the following Lagrangian
$\mathcal{L} =\mathcal{L}_0 + \mathcal{L}_1   +\mathcal{L}_2  $ where
\beqn \mathcal{L}_0 =
    - \frac{1}{4}F_{1\mu\nu}F_1^{\mu\nu}
    - \frac{1}{4}F_{2\mu\nu}F_2^{\mu\nu}
    - \frac{\delta}{2}F_{1\mu\nu}F_2^{\mu\nu},
 ~~~~\mathcal{L}_1 =
     J'_{\mu}A_1^{\mu}
    +J_{\mu}A_2^{\mu},\nonumber\\
    {\cal{L}}_{2} = -\frac{1}{2} M_1^2  A_{1\mu}A_1^{\mu}
-\frac{1}{2} M_2^2 A_{2\mu}A_2^{\mu} - M_1M_2 A_{1\mu}A_2^{\mu}.
        \label{ktst}
\eeqn Here $J_{\mu}$ is the current arising from the physical sector
including quarks, leptons, and the Higgs fields and $J_{\mu}'$ is
the current arising from the hidden sector. As indicated in the
discussion preceding Eq.(\ref{kinetic}), the mass matrix can be
diagonalized by the $R$ transformation which for this $2\times 2$
example is parameterized as follows
\begin{equation}
    R=\left[\matrix{ \cos\theta & -\sin\theta \cr \sin\theta &
    \cos\theta}\right],
\end{equation}
where $\theta$ is determined by the diagonalization constraint  so that
\begin{equation}
\theta=\arctan\left[\frac{\epsilon\sqrt{1-\delta^2}}{1-\delta\epsilon}\right].
\end{equation}
The diagonalization yields one massless mode $A^{\mu}_{\gamma}$ and
one massive mode $A^{\mu}_M$. In this case the interaction
Lagrangian in the diagonal basis assumes the
form\cite{Feldman:2007wj}
\begin{eqnarray}
{\cal{L}}_{1} &= &\frac{1}{\sqrt{1-2\delta\epsilon+\epsilon^2}} \left( \frac{\epsilon-\delta}{\sqrt{1-\delta^2}} J_{\mu} +
       \frac{1-\delta\epsilon}{\sqrt{1-\delta^2}} J_{\mu}' \right) A_M^{\mu}
            \nonumber\\ & + &   \frac{1}{\sqrt{1-2\delta\epsilon+\epsilon^2}}
            \left(J_{\mu}- \epsilon J_{\mu}' \right)  A^{\mu}_{\gamma}.
\end{eqnarray}
\noindent The interesting phenomenon to note here is that the photon
field  $A_{\gamma}^{\mu}$ couples with the hidden sector current
$J_{\mu}'$  only due to mass  mixing, i.e., only due to $\epsilon$.
Thus the origin of milli charge is due to the \st  mass
mixing both in the presence or absence of kinetic mixing. This
phenomenon persists when  one considers $G_{SM}\times U(1)_X$ where
the $SU(2)_L\times U(1)_Y$ gauge group is broken by the conventional
Higgs mechanism and in addition one has the Stueckelberg mechanism
generating a mass mixing between the $U(1)_Y$ and $U(1)_X$.  The
above phenomenon is to be contrasted with the kinetic mixing model
\cite{Holdom:1985ag} where one has two massless modes (the photon
and the paraphoton)  and the photon can couple with the hidden
sector because of kinetic mixing generating milli charge couplings.
[An analysis with kinetic mixing and mass mixings of a different type than discussed here is 
 also considered in \cite{Holdom:1990xp}].

\begin{figure}
  \includegraphics[height=.3\textheight]{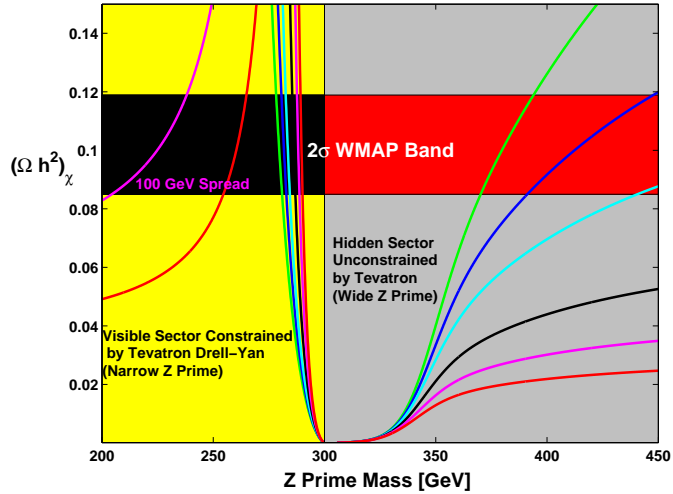}
  \caption{An analysis of the  relic density of milli-charged
particles for the case  when kinetic mixing is included in the \st
$Z'$ model. The analysis is done for $M_{\chi} = $ 150 GeV,
$\bar\epsilon = 04$,
  and  $\delta = (.05, .075, .10, .15, .20, .25)$,
  where the values are in descending order for $M_{Z'}>300$ GeV.
  The red and black bands are the WMAP-3 constraints where the black
  band also produces an observable dilepton signal.
  The analysis shows that for $\bar \epsilon$ fixed,
 increasing $\delta$ increases the parameter space where the WMAP-3 relic density constraint is satisfied,
 while  allowing for  a detectable $Z$ prime signal as shown in Fig.(\ref{fig1}). From  \cite{Feldman:2007wj}.}

\label{fig2}
\end{figure}
\subsection{Milli charge dark matter}
The hidden sector particles are typically natural candidates  for
dark matter. The main issue concerns their ability to annihilate in
sufficient amounts to satisfy the current relic density constraints.
Now the milli charged particles could decay in sufficient amounts by
decaying via the $Z'$  to the Standard Model particles  if their
masses are $<M_{Z'}/2$.  An explicit analysis of this possibility is
carried out in  \cite{Cheung:2007ut} where a pair of Dirac fermions
were put in the hidden sector which couple with strength $g_2$ with
the Stueckelberg field $C_{\mu}$. In this case it was shown that the
relic density constraints consistent with the WMAP-3 data can be
satisfied. Further, with inclusion of proper thermal averaging  of
the quantity $\langle\sigma v\rangle$ over the resonant $Z'$ [using
techniques discussed in
\cite{Nath:1992ty,Baer:1995nc,Gondolo:1990dk,Griest:1990kh,Arnowitt:1993qp}]
  which enters in the relic density analysis, one finds that the WMAP-3
relic density constraints can also be satisfied over a broad range
when the masses of the milli charged hidden sector particles lie
above $M_{Z'}/2$ , with and without kinetic
mixing\cite{Feldman:2007wj}. This phenomenon comes about because of
the thermal averaging effect. On the branch where the milli charged
particles have masses lying above $M_{Z'}/2$ the relic density
constraints can be satisfied and still produce a dilepton signal
which may be observable at the LHC. \cite{Feldman:2007wj}.
Satisfaction of the relic density constraints consistent with WMAP-3
and illustration of the strong dilepton signal are seen in
Figs.(\ref{fig1},\ref{fig2})[taken from \cite{Feldman:2007wj}]. The
experimental constraints on milli charged particles have been
discussed in a number of papers in the literature mostly in the
context of  kinetic mixing models,
\cite{Goldberg:1986nk,Golowich:1986tj,Mohapatra:1990vq,Davidson:1993sj,Foot:1989fh,Caldwell:1988su,Dobroliubov:1989mr,Davidson:2000hf,Perl:2001xi,Prinz:1998ua,Dubovsky:2003yn,Badertscher:2006fm,Gninenko:2006fi},
but without mass generation via the \st mechanism.

%%%%%%%%%%%%%%%%%%%%%%%%%%%%%%%%%%%%%%%%%%%%%%%%
%% BACKMATTER
%%%%%%%%%%%%%%%%%%%%%%%%%%%%%%%%%%%%%%%%%%%%%%%%

\begin{theacknowledgments}
This work was supported in part by the NSF grant PHY-0456568. One of us (PN) 
acknowledges  the  hospitality extended him by Dr. Alok Misra and by other  conference 
organizers at Roorkee.
\end{theacknowledgments}



\bibliographystyle{aipproc}   % if natbib is available

%\bibliography{sample}

%\hyphenation{Post-Script Sprin-ger}

\hyphenation{Post-Script Sprin-ger}

%
%\end{thebibliography}
%\bibliography{sample}
\IfFileExists{\jobname.bbl}{}
 {\typeout{}
  \typeout{******************************************}
  \typeout{** Please run "bibtex \jobname" to optain}
  \typeout{** the bibliography and then re-run LaTeX}
  \typeout{** twice to fix the references!}
  \typeout{******************************************}
  \typeout{}
 }

\end{document}